# Ultrasound-Coupled Microdroplet Laser Chip for High-Throughput Hyperlipidemia Screening

Zhonghao Li[1], Zhihan Cai[1], Chaoyang Gong[1,*]

[1] Key Laboratory of Optoelectronic Technology and Systems (Ministry of Education of China), School of Optoelectronic Engineering, Chongqing University, Chongqing 400044, China.

Correspondence Email: cygong@cqu.edu.cn

**Abstract**

The mechanical properties of biological fluids can serve as early indicators of disease, offering valuable insights into complex physiological and pathological processes. However, the existing technologies can hardly support high throughput measurement, which hinders their broad applications in disease diagnosis. Here, we propose the ultrasound-coupled microdroplet laser chips to enable high-throughput measurement of the intrinsic mechanical properties of fluids. The microdroplets supporting high-Q ($10^4$) whispering gallery modes (WGM) lasing were massively fabricated on a hydrophobic surface with inject printing. The ultrasound was used to actuate the mechanical vibration of the microdroplets. We found that the stimulus-response of the laser emission is strongly dependent on the intrinsic mechanical properties of the liquid, which as subsequently employed to quantify the viscosity. The ultrasound-coupled microdroplet laser chips were used to monitor molecular interactions of bovine serum albumin. High-throughput screening of hyperlipidemia disease was also demonstrated by performing over 2,000 measurements using fast laser scanning. Thanks to the small volume of the microdroplets, a single drop of blood can support over eight billion measurements. The high-throughput ability and small sample consumption of the microlaser chip make it a promising tool for clinical diagnoses based on mechanical properties.

# 1. Introduction

Biological fluid analysis serves as the foundation for diagnostics and clinical decision-making across a broad range of pathologies[1-3]. The mechanical properties of biological fluids, such as viscosity and surface tension, can serve as early indicators of disease, providing rich information in understanding the complex physiological and pathological processes[4-6]. For example, alterations in blood viscosity are associated with cardiovascular diseases and can precede detectable biochemical changes[7,8]. Similarly, the surface tension of fluids such as serum and saliva has been linked to conditions like cancer, where variations may signal changes in tumor microenvironments[9,10]. Techniques based on rheological measurements[11,12], atomic force microscopy (AFM)[13,14], Brillouin spectroscopy[15,16], optical tweezers[17,18], and microfluidics[19,20] have been developed to assess the mechanical properties of biological fluid, driving the rapid expansion of mechanobiology. However, the low throughput of these existing technologies limits their potential for diagnostic applications. Therefore, developing high-throughput methodologies for extracting mechanical information from biological fluids is highly desirable to enhance diagnostic efficiency and expand biomedical applications.

Optical microcavities are recently emerging as a powerful tool to detect mechanical properties in biological systems[21,22]. Among them, the whispering gallery mode (WGM) microcavity has drawn much attention due to its ultrahigh Q-factor up to $10^9$, providing significantly enhanced intracavity light-matter interaction[23-25]. The interaction between optical modes and mechanical stimulus enables precisely quantifying essential mechanical parameters such as force[26], pressure[27], viscosity[28], and vibration frequency[29-31]. In these configurations, near-field excitation of high Q-factor WGMs is required to achieve high sensitivity, which compromises system stability and restricts high-throughput measurements. In contrast, the WGM microlasers, which integrate gain medium within WGM microcavities, enable free-space excitation and signal collection. By leveraging their far-field coupling capabilities, WGM microlasers can serve as free probes that can be dispersed in various biological environments, enabling intracellular[32] and deep tissue force measurements[33]

However, due to their inability to apply stimulus to the external environment, current WGM microlasers struggle to reveal the intrinsic mechanical properties of biological samples.

Here, we propose the concept of an ultrasound-coupled microdroplet laser chip, which utilizes ultrasound to stimulate microdroplets, and the vibrational response was reconstructed with laser emission. As illustrated in Fig. 1a, the dye-filled microdroplet array was fabricated with a commercial inject printer and was excited in free space with a pulsed laser. The spherical microdroplet supports high-Q factor WGM, providing strong optical feedback for lasing. The ultrasound applied to the liquid droplet induces mechanical vibrations, resulting in periodic intensity fluctuations and wavelength shifts of laser emission (Figs. 1b and 1c). Our findings indicate that the vibrational behavior obtained from laser emission is strongly dependent on the intrinsic viscosity of the liquid, offering mechanical insights into molecular interactions (Fig. 1d). Thanks to the easy coupling of microdroplets with the far-field excitation, high-throughput measurements of serum from hyperlipidemia patients were demonstrated by fast laser scanning, indicating its great potential application for clinical diagnostic applications. Our work offers a high-throughput solution for quantifying molecular interactions in liquids, paving the way for clinical diagnoses based on mechanical properties.

## 2. Results
### 2.1 Coupling mechanism of ultrasound and microdroplet laser

An upright microscope system was used for laser excitation and signal collection (Supplementary Fig. S1). The microdroplets with a diameter of 20 μm were massively fabricated on a superhydrophobic surface by a commercial inject printer, forming a contract angle larger than 130° (Supplementary Fig. S2). The microdroplets with nearly spherical morphology support whispering gallery modes with a high Q-factor of up to $6 \times 10^4$ (Supplementary Fig. S3), providing strong optical feedback for lasing. As illustrated in Fig. 2a, due to the tangential radiation of the WGM, bright laser emission started appearing at the edge of the microdroplet after the pump energy density was above the lasing threshold. The observed laser pattern of the liquid droplet can be regarded as the superposition of spatial modes with different lasing wavelengths. To visualize the components of the laser modes, we imaged the laser pattern with a spectrograph system (Fig. 2b). Due to the dispersive nature of the spectrograph system, the laser mode pattern is resolved based on its wavelength (Figure 2c and Supplementary Fig. S4), allowing for the identification of mode components with slight phase differences. Unlike the Laguerre-Gaussian modes observed in Fabry-Perot (FP) cavities[34,35], the laser modes of a WGM cavity consist of two symmetrically crescent-shaped components, the orientation of which is determined by the optical oscillation within the droplet (Fig. 2c). The laser spectrum is shown in Fig. 2d, with individual strong peaks corresponding to the longitudinal laser modes. Because of the high Q-factor of the liquid droplet, a low lasing threshold of about 10.5 μJ/mm$^2$ was obtained (Fig. 2e), which is comparable to the other microdroplet lasers[36].

Then, we investigated the coupling mechanism of ultrasound and microdroplet lasers by monitoring the temporal evolution of spectral images. As shown in Fig. 2f, the spectral images at different observation times exhibited significantly different orientations, showing a period of approximately 0.2 seconds (More details about the period changes of laser emission will be explained in the following section). In contrast, the spectral images remain stable in the absence of ultrasound (Fig. 2g). This result can be explained by the periodic mechanical vibration actuated by ultrasound. As illustrated in Fig. 1b, the ultrasound exerted a time-varying force on the microdroplet, inducing a periodic deformation in morphology. When the microdroplet changes from spherical to elliptical, the Q-factor of the WGM drops due to chaotic resonance[37]. As a result, only WGMs with higher Q-factors support lasing in mode competition. The divergence of the oscillation direction reveals the asymmetrical geometry of the microcavity, which offers a potential approach for measuring the anisotropy of mechanical forces in the surrounding environment[38].

**2.2 Quantification of mechanical vibration via laser spectrum**
The morphological changes in microdroplet alter the periodic switching of WGMs oscillation direction, resulting in a significant wavelength shift of laser peaks (Fig. 3a). Figure 3b shows the periodic wavelength shift of the laser peak at 617.3 nm, with a maximum shift of approximately 0.08 nm, corresponding to a small geometric deformation of 6.5 nm. This value is below the diffraction limit of the microscope, which confirms the extraordinary ability of the WGM laser in measuring slight mechanical vibrations. Furthermore, due to the varying intracavity losses encountered by the WGMs in different orientations, intensity fluctuation in the lasing peak was also observed (Fig. 3c). Interestingly, distinct wavelength shifts and intensity variations are observed in WGMs with different orders (Figs. 3d and 3e), which is due to the spatial heterogeneity of WGMs on the microdroplet. This relatively large divergence hinders the accurate quantification of the mechanical vibration by routinely tracking the wavelength shift or intensity fluctuation[39].

To overcome this issue, we calculated the temporal correlation of laser spectra, incorporating the lasing wavelength shifts and intensity variations in the analysis (see Methods for details). As illustrated in Fig. 3f, the temporal correlation curve exhibits a fast fluctuation superimposed on a slow envelope. The fast Fourier transform (FFT) spectrum has a sharp peak at $f_1$ = 0.3 Hz, which shows an identical peak location and a higher signal-to-noise ratio compared to the FFT spectra of wavelength shifts and intensity variations (Supplementary Fig. S7). This result indicates that the temporal correlation accurately captures the features of the laser spectra evolution. We have to note that the peak frequency observed in the FFT spectrum is five orders of magnitude lower than the ultrasound excitation (132.4 kHz). This is owing to the undersampling effect caused by the low pump repetition rate (20 Hz) (Supplementary Figs. S7 to S9). To quantify the mechanical vibration of the microdroplet, the standard deviation (SD) of the correlation curve was used as the output signal, which reflects the deformability under ultrasound excitation and is independent of the pulsed laser repetition rates (Fig. 3g).

## 2.3 Monitoring molecular interaction of bovine serum albumin

Subsequently, ultrasound-coupled microdroplet lasers were employed to measure the mechanical properties of liquid. Viscosity was used as an example since it is strongly related to the molecular interaction in solution. As shown in Fig. 4a, a liquid's viscosity originates from intermolecular forces that resist its flow. We fabricated microdroplet microlasers with various viscosities by varying the concentration of glycerol

(Supplementary Fig. S10) and investigated their vibrational response to ultrasound stimulation. As shown in Fig. 4b, the viscosity of the liquid droplet has a significant impact on their vibration behavior. Because the stronger intermolecular forces in high viscosity liquid inhibit the mechanical vibration, a narrower statistical distribution was observed. We calibrated the SD of the temporal correlation curve with various viscosities in Fig. 4c, showing a nonlinear response curve with maximum sensitivity in the linear range (2.8 to 10 mPa.s). This result indicates that the microdroplet lasers are appropriate for monitoring subtle mechanical changes in low viscosity liquid.

Bovine serum albumin (BSA) was chosen as the model protein to investigate molecular interactions under different conditions. BSA has an isoelectric point of about pH = 5.1 and carries a negative net charge under neutral conditions. As a result, the BSA molecules experience electrostatic repulsion forces from the nearby molecules, which suppresses the mobility of BSA molecules. As the BSA concentration increases, the electrostatic repulsion forces increase and lead to a higher viscosity. As illustrated in Fig. 4d, the viscosity of BSA solution shows a linear relationship with concentration in the range from 50 to 150 mg/ml. When NaCl was inducted into the BSA solution, the strong charge shielding effect of $Na^+$ and $Cl^-$ weakened the electrostatic forces between BSA molecules[40]. Therefore, increasing the NaCl concentration led to a decrease in the viscosity of the BSA solution (Fig. 4e). As the pH decreases from neutral to acidic conditions, the viscosity of the BSA solution decreases, which can be attributed to protein denaturation under acidic conditions (Fig. 4f)[41].

## 2.3 High-throughput screening of hyperlipidemia

Hyperlipidemia is caused by high levels of fats in the blood and affects over 25 million people worldwide. It is diagnosed through elevated cholesterol levels, which increase blood viscosity and can lead to cardiovascular events such as strokes and heart disease[42,43]. The capability of the microdroplet lasers to measure liquid viscosity indicates their significant potential for diagnosing hyperlipidemia.

Massive production of serum laser chips were achieved by inject printing technology (Fig. 5a) (See Methods for more detailed information). The location and size of each droplet can be precisely controlled with the printer, forming a programmable microlaser array (Figs. 5b and 5c). Laser scanning was employed to achieve high-throughput measurement of the mechanical vibration of each droplet. In order to achieve a relatively fast scanning speed, a pulsed laser with a 500 Hz repetition rate was used for laser excitation. As shown in Fig. 5d, the temporal evolution correlation curve contains the vibration signal of each droplet. Due to the precise control of droplet size during inject-printing, the performance of various microdroplet lasers exhibits excellent consistency, with a low variation of 1.8% (Supplementary Fig. 11).

To demonstrate its potential application in clinical diagnosis, we prepared four serum laser chips fabricated with samples from four patients. Fig. 5e shows the viscosity of each measurement, indicating four stages (normal, stage 1, stage 2, and stage 3) with increasing cholesterol levels (Supplementary Table 1). The greater divergence of the results at higher cholesterol levels arises from the nonlinear nature of the viscosity curve in Fig. 4c. The relative low slope of the curve at higher viscosities results in greater errors in viscosity measurements. In this experiment, viscosities of over 2000 droplets were extracted within 25 min. The scanning speed can be further improved by increasing the pump repetition rate. Thanks to the small volume of the microdroplets ($4.18 \times 10^{-6}$ μL), a single drop of blood (~ 35 μL) can support over eight billion measurements. The high-throughput ability and small sample consumption of the microlaser chip make it a promising tool for clinic diagnosis.

## 3. Discussion

We proposed the ultrasound-coupled microdroplet laser chip to realize high-throughput detection of liquid viscosity. The ultrasound actuated the mechanical vibration of microdroplets, resulting in periodic intensity fluctuations and wavelength shifts in the lasing spectra, which were utilized to assess the droplets' stimulus-response to ultrasound. We investigated the coupling mechanism between ultrasound and laser modes. We found that the stimulus-response of laser emission is highly dependent on liquid viscosity. The molecular interaction of BSA molecules and the high-throughput screening of hyperlipidemia were demonstrated by monitoring the ultrasound-induced vibration. Our findings pave the way for high-throughput mechanical analysis of liquid biological samples, which can potentially be employed for low-cost clinical diagnosis.

The key idea of this work is to detect the response of microdroplets to ultrasound stimulation using WGM lasers. Although current optical microcavities play an important role in assessing weak mechanical forces in biological samples, most of them lack the ability to actively apply stimulus to the environment, making it difficult to detect the intrinsic mechanical properties of the biological samples. The proposed stimulus-response measurement of microdroplet lasers provides new mechanical insights into understanding the complex physiological processes. We anticipate that this work will pave the way for applications such as information encryption[44] and the quantification of in vivo mechanical properties[45].

## 4. Methods

### 4.1 Optical system setup

The experimental setup is illustrated in Supplementary Fig. S1. An upright microscope system equipped with a ×50 objective was used for microdroplet laser excitation and signal collection. An optical parametric oscillator (OPOTEK, MagicPRISM VIS), pumped by the third-harmonic wave of a Nd: YAG laser (Beamtech Optronics Co., Ltd., DAWA-200), served as the pump laser. The 532 nm pulsed pump laser (20 Hz repetition rate) was coupled into the upright microscope system and focused to a laser spot with a diameter of 4 μm. The emitted laser from the microdroplet was send to a scientific complementary metal-oxide-semiconductor (sCMOS) camera (Tucsen, Dhyana 400BSI V3) and a spectrometer (Princeton Instruments, SpectrPro-500i) for image and spectral acquisition, respectively. The pump laser, sCMOS camera, and spectrometer were synchronized throughout the experiment to ensure that each spectrum and image corresponded to a single pump pulse. A three-dimensional motorized stage was employed for laser scanning and high-throughput measurements.

### 4.2 Fabrication of microdroplet laser chip

A Rhodamine solution with a concentration of 10 mM was prepared by dissolving Rhodamine B powder (Aladdin, No. 81-88-9) in deionized (DI) water. Subsequently, Rhodamine solution and glycerol (Aladdin, No. 56-81-5) were mixed with a volume ratio of 1:1, and the resulting mixture was used as ink for an inkjet printer. The addition of glycerol to the microdroplet reduces evaporation.

The microdroplet laser chip was fabricated on a glass slide. First, the glass slide was cleaned using a plasma cleaner (PLUTOVAC, PLUTO-T) for 1 min. The cleaned slide was then immersed in a coating solution (fluorinated silica solution) to form a thin superhydrophobic layer. A commercial inkjet printer (EPSON, L130) was used for microdroplet generation, producing microdroplets with a nearly spherical morphology and a large contact angle exceeding 130° (Supplementary Fig. S2). A piezoelectric transducer was adhered to the frontside of the glass slide to induce ultrasound. A sine wave generated by function generator at fixed frequency of 132.4 kHz was used to drive the piezoelectric transducer.

### 4.3 Hyperspectral imaging

Hyperspectral images were recorded using the spectrometer. The laser signal was coupled into the spectrometer through the entrance slit and dispersed by a grating (600 lines/mm) according to wavelength, allowing different spectral components to be distinguished at various locations on the camera. The entrance slit was opened wide enough to collect the entire laser pattern.

### 4.4 Detecting molecular interaction in BSA solution

The BSA stock solution was prepared by dissolving BSA powder (Aladdin, No. B265994) in phosphate-buffered saline (PBS). For the measurements in Fig. 4d, BSA solutions with concentrations ranging from 50 to 150 mg/mL (50, 75, 100, 125, and 150 mg/mL)were prepared by diluting the stock solution with PBS. For the measurements in Fig. 4e, NaCl solutions with varying concentrations (0, 150, 300, 450, and 600 mM) were mixed with the BSA solution with a concentration of 200 mg/mL at a volume ratio of 1:1. For the measurements in Fig. 4f, a BSA solution with a concentration of 100 mg/mL was obtained by diluting the stock solution with deionized (DI) water.

Subsequently, BSA solutions with different pH values (1, 3, 5, and 7) were prepared by adjusting the pH with acetate (Aladdin, No. A116166) and NaOH (Aladdin, No. S431790) solutions.

**4.5 Preparing of serum**

Human blood from different patients was freshly collected in Chongqing General Hospital. Then, the blood samples were centrifugated at 4000 r/min for 10 minutes. The resulting supernatant was transferred to a clean microcentrifuge tube and stored at -80 ℃ for future use. In high-throughput laser scanning experiment, a 10 times dilution with PBS of serum was mixed with Rhodamine B (50 mM) and glycerol (volume: 50%) in PBS with a volume ratio of 18:1:1. The mixed solution was further injected into the printer for microdroplets fabrication.

**Acknowledgments**

This work is supported by the National Natural Science Foundation of China (Grant No. 62375030); Fundamental Research Funds for the Central Universities (Grant No. 2024CDJYXTD-004); Young Elite Scientists Sponsorship Program by CAST (Grant No. 2022QNRC001).

**Conflict of interest**

All the authors declare no conflict of interests.

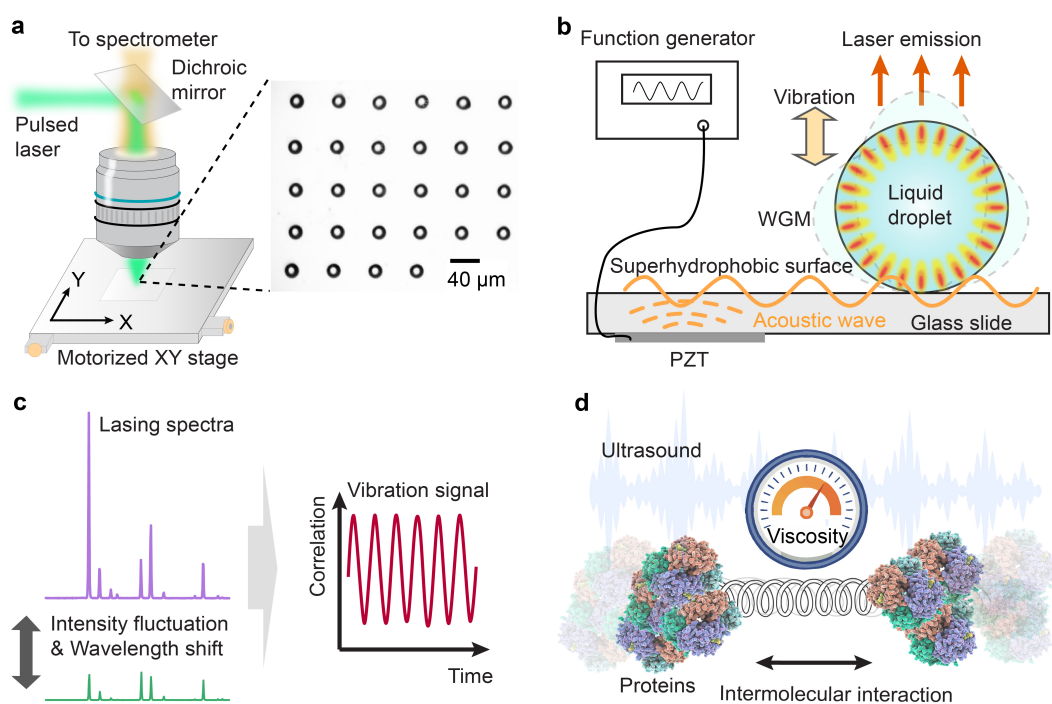

**Figure 1 The principle of ultrasound-coupled microdroplet laser chip.** (**a**) Illustration of the high-throughput laser scanning of the microdroplet laser array. (**b**) Schematic illustration of the coupling between ultrasound and WGMs. (**c**) Extraction of vibration signal from laser spectra. (**d**) Illustration of revealing molecular interactions with ultrasound.

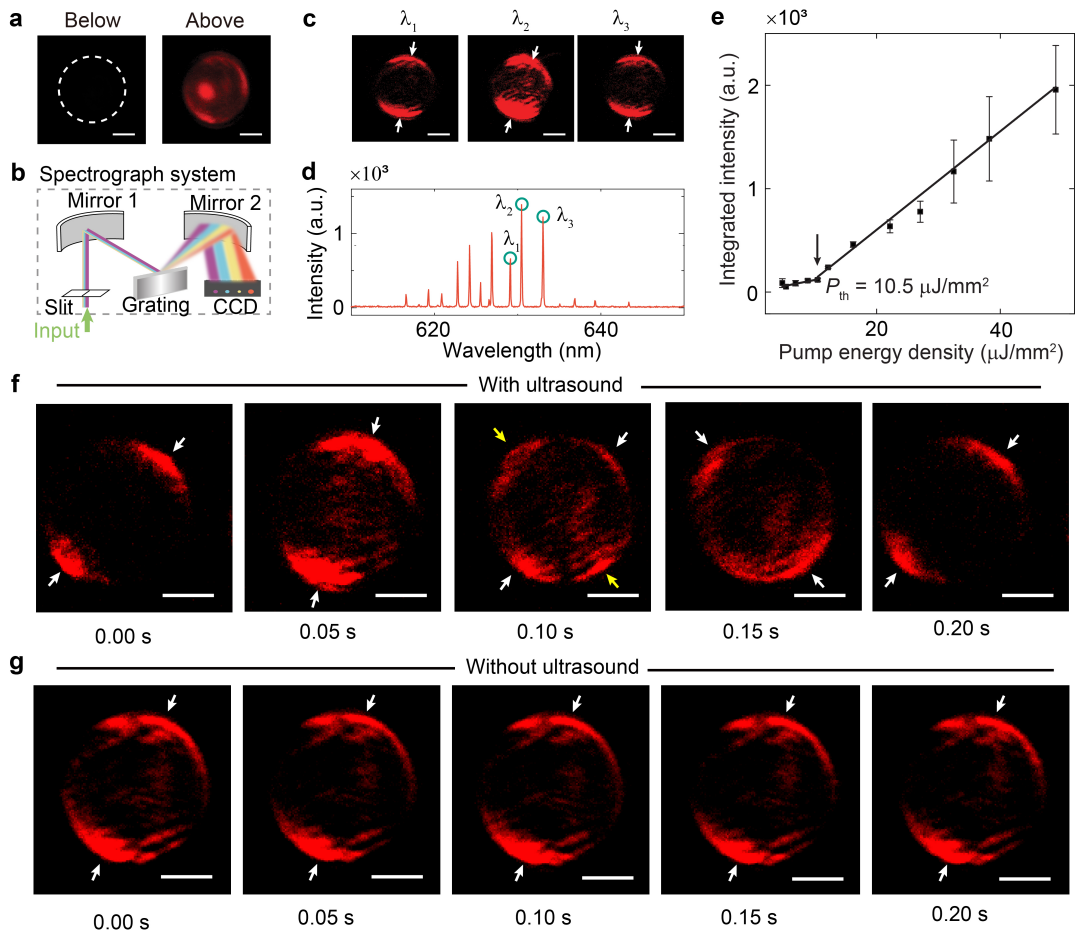

**Figure 2 Coupling mechanism between ultrasound and microdroplet laser.** (**a**) Image of the microdroplet with the pump energy density below (left) and above (right) the lasing threshold. The dashed circles indicate the boundary of the microdroplet. Scale bar: 5 μm. (**b**) Illustration of the spectrograph system. The laser pattern was dispersed by a diffractive grating according to its spectral components. (**c**) Spectral components of the laser pattern in (**a**). Scale bar: 5 μm. (**d**) Laser emission spectrum of the droplet in (**a**). (**e**) The spectrally integrated intensity as a function of various pump energy densities. (**f, g**) Temporal evolution of laser spectral images with (**f**) and without (**g**) ultrasound excitation. The arrows indicate the orientations of the WGMs. Images are recorded at 629.1 nm. Scale bar: 5 μm

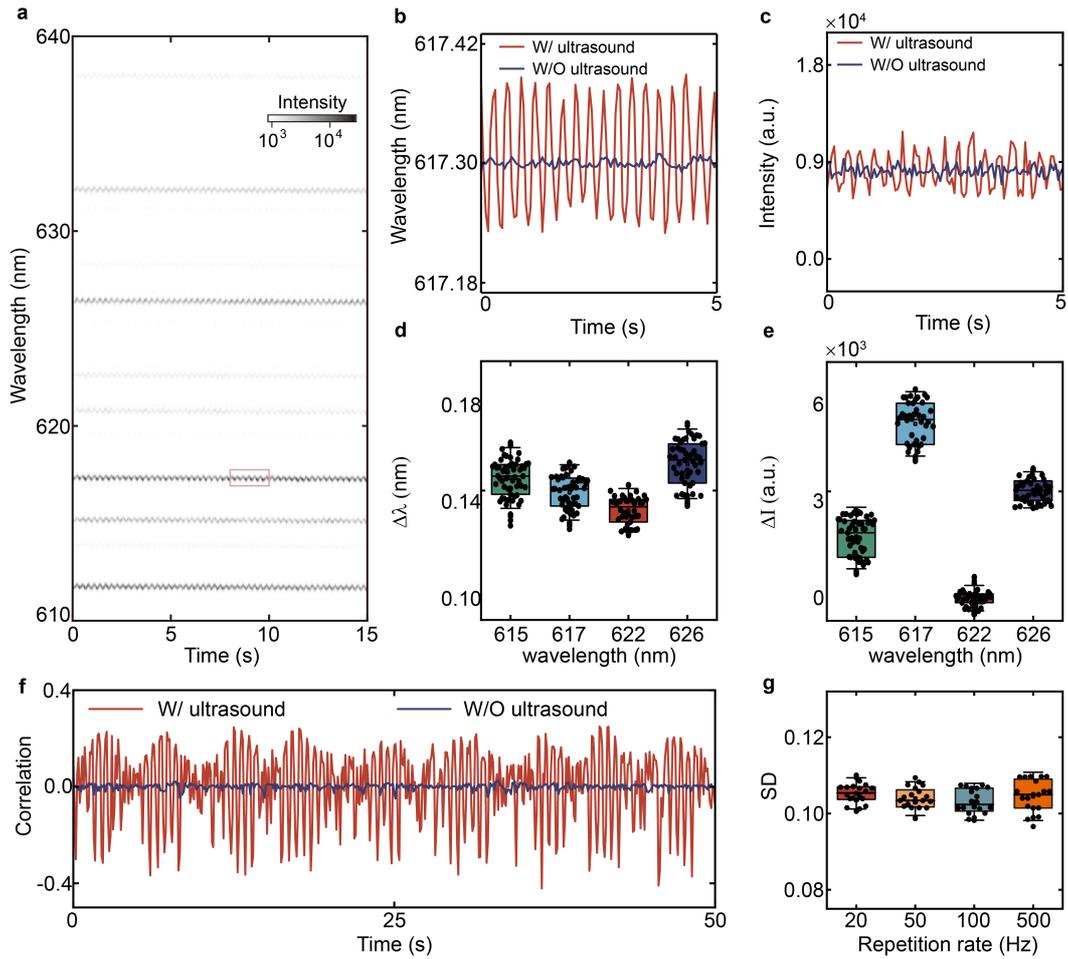

**Figure 3 Quantification of mechanical vibration with laser emission.** (**a**) Temporal evolution of lasing spectra induced by ultrasound. (**b, c**) Temporal evolution of wavelength and intensity of the lasing peak. Data are extracted from the boxed region in (**a**). (**d,e**) Wavelength shifts (**d**) and intensity fluctuations (**e**) obtained from different orders of lasing modes. (**f**) Comparison of time-varying correlation curves with (red) and without (blue) ultrasound. (**g**) Standard division of the correlation curves with different pump repetition rates.

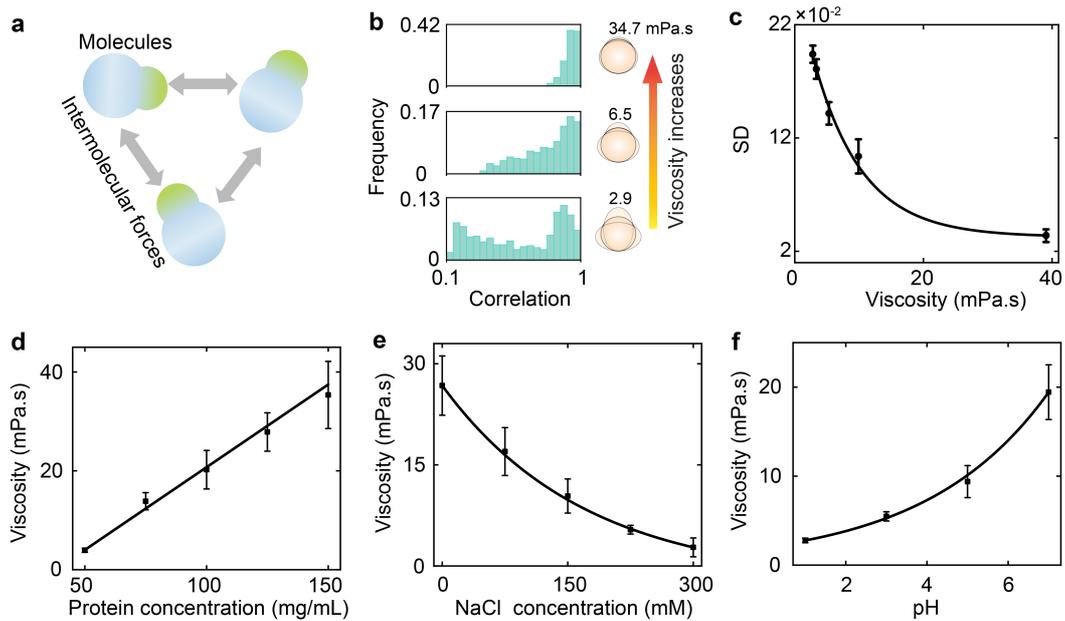

**Figure 4 Probing molecular interaction of BSA protein.** (**a**) Schematic illustration of molecular interaction. (**b**) Statistics results of the correlation with various glycerol concentrations. (**c**) Calibration curve. (**d,e,f**) Viscosity as a function of protein concentration (**d**), NaCl concentration (**e**) and pH (**f**).

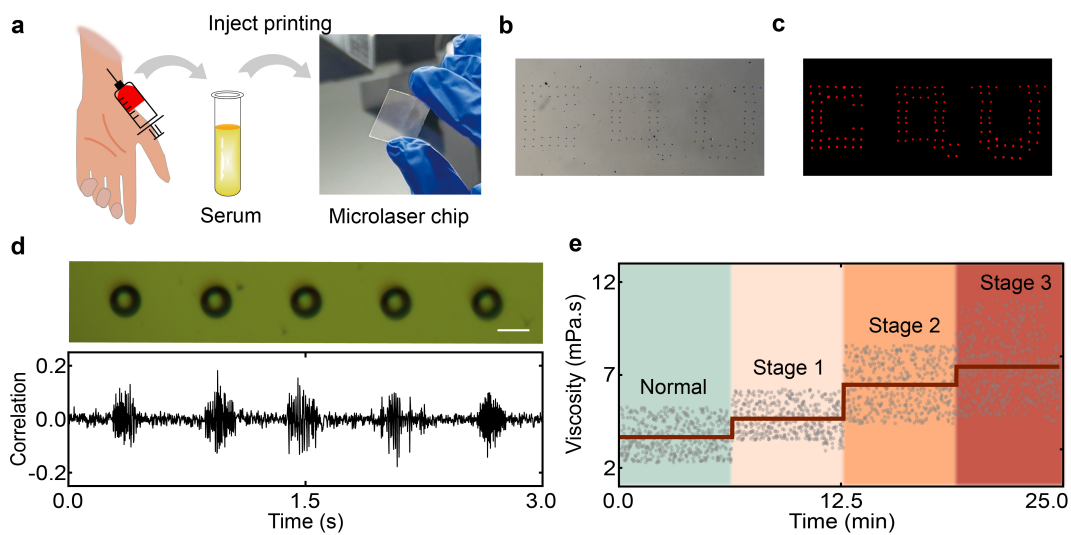

**Figure 5 High-throughput measurement of serum.** (**a**) Illustration of fabricating microdroplet laser chip using human serum. (**b,c**) Comparison of bright filed (**b**) and fluorescence image (**c**) of the droplet array forming the character "CQU". (**d**) Time-varying correlation (**bottom**) of five microdroplets (**top**). Scale bar: 20 μm (**e**) Viscosity measurement result of serum.